\documentclass[aps,reprint,superscriptaddress,floatfix]{revtex4-1}

\usepackage{hyperref}
\usepackage{amsmath}
\usepackage{amssymb}
\usepackage{xspace}
\usepackage{graphicx}

\newcommand{\ie}{i.e.\xspace}
\newcommand{\eg}{e.g.\xspace}
\newcommand{\Acm}{\ensuremath{\mathrm{A} / \mathrm{cm}^{2}}\xspace}
\newcommand{\mum}{\ensuremath{\mu \mathrm{m}}\xspace}

\begin{document}

\title{X-ray spectroscopy of current-induced spin-orbit torques and spin accumulation in Pt/3\textit{d} transition metal bilayers}

\author{C.~Stamm} \email[]{christian.stamm@mat.ethz.ch}
\author{C.~Murer}
\affiliation{Department of Materials, ETH Z{\"u}rich, 8093 Z{\"u}rich, Switzerland}

\author{Y.~Acremann}
\affiliation{Laboratory for Solid State Physics, ETH Z{\"u}rich, 8093 Z{\"u}rich, Switzerland}

\author{M.~Baumgartner}
\affiliation{Department of Materials, ETH Z{\"u}rich, 8093 Z{\"u}rich, Switzerland}

\author{R.~Gort}
\author{S.~D{\"a}ster}
\affiliation{Laboratory for Solid State Physics, ETH Z{\"u}rich, 8093 Z{\"u}rich, Switzerland}

\author{A.~Kleibert}
\affiliation{Swiss Light Source, Paul Scherrer Institut, 5232 Villigen PSI, Switzerland}

\author{K.~Garello}
\author{J.~Feng}
\affiliation{Department of Materials, ETH Z{\"u}rich, 8093 Z{\"u}rich, Switzerland}

\author{M.~Gabureac}
\affiliation{Department of Materials, ETH Z{\"u}rich, 8093 Z{\"u}rich, Switzerland}
\affiliation{Laboratory for Solid State Physics, ETH Z{\"u}rich, 8093 Z{\"u}rich, Switzerland}

\author{Z.~Chen}
\author{J.~St{\"o}hr}
\affiliation{Stanford Institute for Materials and Energy Sciences, SLAC National Accelerator Laboratory, 2575 Sand Hill Road, Menlo Park, California 94025, USA}

\author{P.~Gambardella}
\affiliation{Department of Materials, ETH Z{\"u}rich, 8093 Z{\"u}rich, Switzerland}

\date{12 July 2019}

\begin{abstract}
An electric current flowing in Pt, a material with strong spin-orbit coupling, leads to spins accumulating at the interfaces by virtue of the spin Hall effect and interfacial charge-spin conversion.
We measure the influence of these interfacial magnetic moments onto adjacent 3\textit{d} transition metal layers by x-ray absorption spectroscopy and x-ray magnetic circular dichroism in a quantitative and element-selective way, with sensitivity below $10^{-5}~\mu_B$ per atom. 
In Pt(6~nm)/Co(2.5~nm), the accumulated spins cause a deviation of the Co magnetization direction, which corresponds to an effective spin-Hall angle of 0.08. 
The spin and orbital magnetic moments of Co are affected in equal proportion by the absorption of the spin current, showing that the transfer of orbital momentum from the recently predicted orbital Hall effect is either below our detection limit, or not directed to the 3\textit{d} states of Co. 
For Pt/NM (NM = Ti, Cr, Cu), we find upper limits for the amount of injected spins corresponding to about $3\times 10^{-6}~\mu_B$ per atom.
\end{abstract}

\maketitle

\section{Introduction}

The coupling of electron spin and charge transport encompasses a broad range of fundamental issues in condensed-matter physics, with straightforward implications for the development of magnetic storage and sensing devices.
Spin currents, which mediate the transfer of angular momentum from one material (or part of it) to another, play a central role in this field, as they permit one to manipulate the magnetization, electrical resistance, and heat flow in both metallic and insulating systems \cite{BookSpinCurrent}.
Pure spin currents can be generated via a spin-polarized charge current injection from ferromagnets, spin Hall effect, interfacial charge-spin conversion, spin pumping, or thermal gradients.
In most experiments, spin currents are detected electrically \cite{Valenzuela2006,Niimi2015}, optically \cite{Kato2004,Stamm2017}, in spin pumping measurements \cite{Saitoh2006}, or through their influence on the magnetization via spin-orbit torques \cite{Garello2013,Kim2013,Avci2014,Baumgartner2017}.
Recently, a first direct x-ray spectroscopic detection of spin currents was realized in scanning transmission x-ray microscopy by performing AC (current on/off) x-ray magnetic circular dichroism (XMCD) measurements at the Cu $L_3$ edge in a Co/Cu nanometer-sized pillar \cite{Kukreja2015}.
This measurement revealed a current-induced magnetic signal in the nonmagnetic Cu layer arising from the spin polarized current through the heterostructure.
The small transient spin signal exhibits a peak at the Fermi level, ascribed to the conduction electrons, and another one that coincides with the static spin signal from Cu in proximity to Co \cite{Samant1994}.
These results motivate x-ray spectroscopy investigations of spin currents and their action in systems with high spin-orbit coupling, notably heavy metal/normal metal heterostructures.

Here we report x-ray absorption spectroscopy (XAS) and XMCD measurements of the current-induced spin accumulation in Pt/3\textit{d} transition metal bilayers. In Pt/Co samples we detect the accumulated spins through the presence of the spin-orbit torque acting on the magnetic Co layer. The Co spins, initially aligned in the film plane, are slightly tilted out of the plane by an amount proportional to the injection current density. Thereby the ratio of the spin to orbital moment in Co remains unchanged.  We further investigated the ability of the spin accumulation to expand from Pt into adjacent layers of the nonmagnetic transition metals Ti, Cr, and Cu.
Here we find that any current-induced XMCD signal at the $L_3$, $L_2$ absorption edges of the 3\textit{d} metal is comparable in size or smaller than our detection limit, which  allows us to give an upper limit for the spin current flowing from Pt into the nonmagnetic layer.

\section{Experimental Procedure}

\begin{figure}[tb]
	\includegraphics[width=8.5cm]{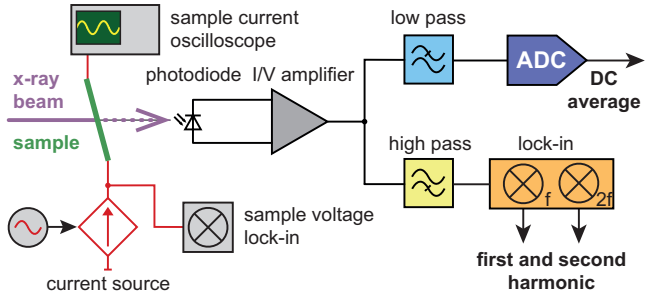}
	\caption{Schematic drawing of the x-ray transmission setup: it comprises the sample, an in-vacuum photodiode, amplifier, low/high pass filter, ADC, and lock-in amplifier. A sine wave drives the current source, which generates the current that is passed through the sample, monitored by a second lock-in and oscilloscope.}
	\label{setup}
\end{figure}

The experiments were performed at the Surface/Interface: Microscopy (SIM) beamline of the Swiss Light Source at the Paul Scherrer Institut.
We measured the absorption of soft x-rays in a thin film sample by recording the transmitted intensity with a photo-diode.
This technique allows for highly sensitive XAS and XMCD measurements that are not disturbed by the electric and magnetic fields arising from the injected current used to generate spin currents.
After amplification, we split the signal into DC and AC components using a combined low pass/high pass filter.
The AC signal is processed in a lock-in amplifier which measures the response of the sample in phase with the injected sinusoidal current.
A harmonic analysis allows us to separate two effects connected to the first and second harmonic, recorded simultaneously by the lock-in: changes that are directly related to the current are detected in the first harmonic, whereas the second harmonic relates to effects connected to the current squared, \eg the dissipated power in the sample.
The current source modulation frequency was set to 50~kHz for all measurements with the exception of Pt/Cu, for which we used 2030~Hz.
The DC response of the photodiode corresponds to the average transmitted intensity and is used to obtain the time-averaged absorption of the sample, resulting in XAS and XMCD spectra.
To achieve a reasonably good signal to noise ratio in the spectra, multiple photon energy scans were measured and averaged, with the acquisition paused during injection of electrons into the storage ring operating in ``top-up'' mode.
A second lock-in amplifier recorded the applied sinusoidal voltage, and the current through the sample was monitored on an oscilloscope.
During the measurements, the vacuum chamber was filled with 100~mbar of He gas in order to dissipate heat from the current carrying sample.
For an illustration of the experimental setup see Fig.~\ref{setup}.

The samples are bilayers consisting of Pt/3\textit{d} transition metal of either Ti, Cr, Co, or Cu.
They are sputter deposited on top of 200~nm thick Si$_3$N$_4$ membranes of 60~\mum $\times$ 500~\mum size.
A cap layer of 3~nm AlO$_x$ is used to protect the sample from air.
The Pt layers are polycrystalline with (111) texture.
Transmission Kikuchi diffraction measurements show that the average grain size is about 15~nm. 
The rms roughness measured by atomic force microscopy on Pt(5~nm)/Co(1~nm)/AlO$_x$ is about 0.5~nm and depends on the thickness of the Al layer, whereas the rms roughness of a 15~nm Pt layer deposited on Si$_3$N$_4$ is about 0.3~nm. Despite the low roughness, we cannot exclude minor interdiffusion of 3\textit{d} elements in Pt limited to the topmost atomic layers (see, e.g., \cite{Gambardella2000}).
The bilayers were directly deposited onto the Si$_3$N$_4$ membrane, with few nm film thickness as specified in the results section for each sample.
On the back of the Si$_3$N$_4$ membrane a 150~nm thick Aluminium layer is deposited which helps to dissipate the heat from the current during injection.
One Si chip carries four of these membranes, and current strips of 100 or 200~\mum  width are patterned across each one by photo-lithography.
Bonding wires and metallic contact pads are used to electrically connect the current strips, and an external switch box is used to select which structure is subjected to current injection.

\begin{figure}[tb]
	\includegraphics[width=8.5cm]{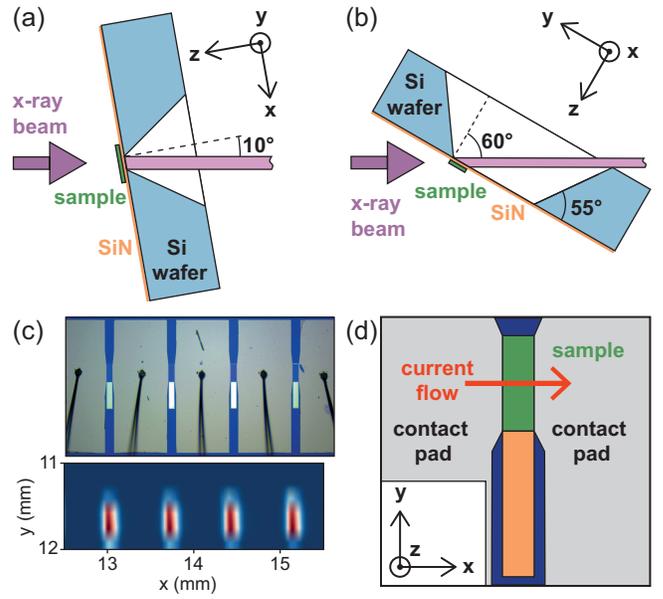}
	\caption{Sample and measurement geometries for (a) close to perpendicular incidence ($10^\circ$) mostly sensitive to the perpendicular magnetization component and (b) tilted ($\approx 60^\circ$) x-ray incidence for detecting in-plane magnetization along $y$.
	(c) Optical micrograph (above) and x-ray transmitted signal recorded by sample scanning (below) of a sample chip with four current strips. Five wires are attached to the metallic contact pads, in between are the Si$_3$N$_4$ membranes with the deposited samples.
	(d) Closeup sketch of one current strip of 200~nm width (green), on top of the 500~nm wide membrane  that extends below the sample (orange). The current flow between the contact pads is indicated by the red arrow.}
	\label{sample}
\end{figure}

Two measurement geometries are employed that have different sensitivities for the spatial magnetization components detected by XMCD.
In Fig.~\ref{sample}(a), the tilt angle is $10^\circ$ and the measurement is mostly sensitive to the out-of-plane magnetization component $M_z$, with a small in-plane $M_x$ contribution.
This geometry is used when measuring the Pt/Co samples with a ferromagnetic Co film.
For the second geometry, sketched in Fig.~\ref{sample}(b), the sample is rotated around $x$ until the x-ray beam passing through the strip is almost cut off by the Si frame underneath the membrane.
Note that from the membrane fabrication process we have a $55^\circ$ cutout angle as a result of the anisotropic etching in Si.
In our geometry, the resulting tilt angle is $\approx 60^\circ$, which makes the XMCD measurement predominantly sensitive to the in-plane $M_y$ component transverse to the current direction, with contributions from $M_z$.
The actual tilt angle is adjusted for each sample in $1^\circ$ steps such that x-rays that did not pass the current strip will not reach the detector. 
This procedure is needed as the FWHM x-ray beam size was set to 100~\mum (horizontal) $\times$ 250~\mum (vertical) \cite{Olivieri2015}, which is slightly larger than our current strip sample.
The tilted geometry of Fig.~\ref{sample}(b) is used in the search for a spin accumulation signal in Pt/nonmagnetic transition metal samples.
The coordinate system, sketched in Fig.~\ref{sample}, is fixed with the sample that has the current line parallel to $x$ and the surface normal along $z$.

For the successful quantitative evaluation of the XAS data, the correct normalization of the signals is necessary.
First, we divide the photodiode signal by the initial intensity as measured on a refocusing mirror, and by the average intensity in the pre-edge region (photon energy $E < E_{L_3}$).
The x-ray absorption, denoted XAS in our graphs, is then determined by taking the negative logarithm of the normalized transmitted intensity, 
$-\ln (I_\mathrm{t}/I_0)$.
The DC average intensity and the lock-in AC components are measured simultaneously and at the same scale, and the above mentioned normalization routine is applied in the same way to all signals, such that they are directly comparable.
The static absorption spectrum is corrected for a slope by subtracting a line fitted in the pre-edge region, which however does not change the scale of the measurement signals.

\section{Results on P\lowercase{t}/C\lowercase{o}}

We begin with describing the measurements on the magnetic bilayer sample Pt(6 nm)/Co(2.5 nm).
A sinusoidal current runs along $x$ through the sample patterned as a strip of 200~\mum width and 70~\mum length, at current densities ranging from $0.79$ to $3.78 \times 10^6$~\Acm, assuming a homogeneous current distribution throughout the Pt/Co structure.
As discussed in the appendix, if one considers the higher resistivity of Pt compared to Co, the actual current density in Pt  is reduced to $\approx 80\%$ of the values stated here.

\begin{figure}[tb]
	\includegraphics[scale=0.53]{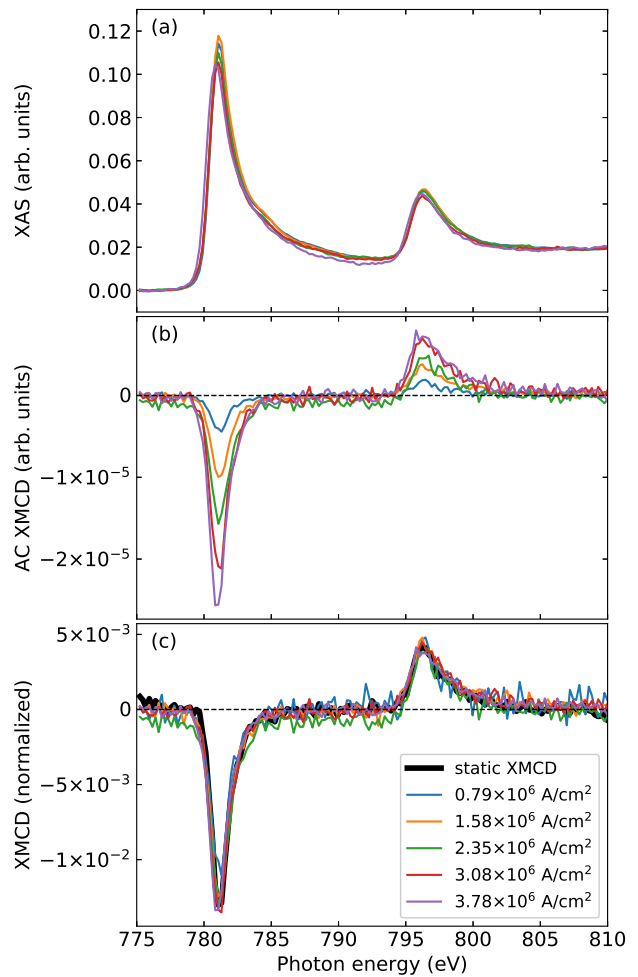}
	\caption{Co spectra of the Pt/Co sample under the influence of current flow, for given current densities:
		(a) average XAS,
		(b) AC-XMCD, and
		(c) AC-XMCD from above rescaled to static XMCD (thick black line).
	}
	\label{currentflow}
\end{figure}

The results of the XAS and XMCD measurements at the Co $L_3$, $L_2$ absorption edges upon injection at various current densities are displayed in Fig.~\ref{currentflow}.
For Pt/Co we use the measurement geometry drawn in Fig.~\ref{sample}(a) with an almost perpendicular x-ray incidence.
A static magnetic field of 42~mT is applied in-plane along the $x$-direction during the measurements, which is sufficient to completely saturate the film and defines the equilibrium magnetization direction.
The XMCD measurement is sensitive to the magnetization vector projected onto the x-ray propagation direction, \ie to both the static in-plane and any out-of-plane component that may arise when current is injected.
The signal from the static XMCD component $M_x$ results from the projection $\sin(10^\circ)=0.17$ onto the beam direction and is seen in the DC average, and is plotted as thick black line in Fig.~\ref{currentflow}(c).
Our measurement is however mostly sensitive to $M_z$ with a projection factor $\cos(10^\circ)=0.98$.
The out-of-plane $M_z$ component is measured in the lock-in first harmonic output and is increasing with the injected current, see Fig.~\ref{currentflow}(b). 
The mechanism behind this behavior is the generation of spin-orbit torques that leads to a change of the equilibrium direction of the magnetization vector $\vec{M}$.
In general, spin-orbit torques are classified into two types, according to their direction with respect to the current flow and the sample magnetization: fieldlike and dampinglike torque \cite{Garello2013,Kim2013,Haney2013}.
Whereas the fieldlike torque leads to a deviation of $\vec{M}$ within the $xy$ plane of the layer, the dampinglike torque can be expressed as a field $B_{DL} \parallel z$, effectively pulling the magnetization vector out of the plane.

In Fig.~\ref{currentflow}(c) we plot the AC-XMCD curves from individual $j$ values, rescaled onto the static XMCD curve.
We note that the curves all have the same shape and follow the static one, demonstrating a constant ratio of $L_3$ to $L_2$ XMCD.
This indicates, as will be confirmed in the sum rule analysis below, an unchanged $m_L/m_S$ ratio of orbital to spin moments in Co.
In the recent discussion on the orbital Hall effect (OHE) \cite{Kontani2009,Go2018, Jo2018} and the orbital Rashba effect \cite{Chen2018}, an orbital momentum is predicted to accompany the spin Hall effect.
The OHE would lead to the accumulation of orbital angular momentum \cite{Go2018}, possibly detectable using XMCD.
Our data however does not show an increased $m_L / m_S$ ratio, at least in the Co 3\textit{d} moments at the interface with Pt.

\begin{figure}[tb]
	\includegraphics[scale=0.53]{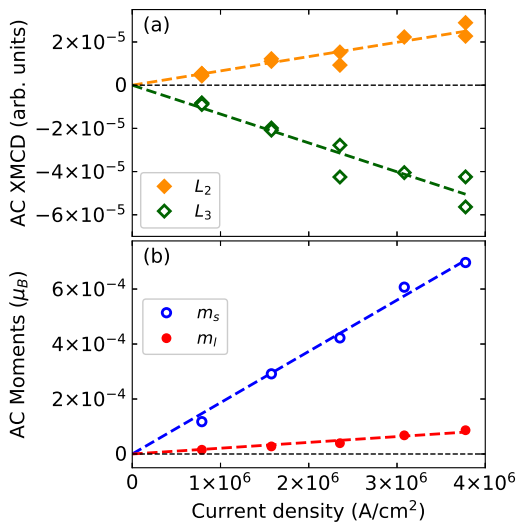}
	\caption{AC-XMCD versus injected current.
		(a) Integrated around the Co $L_2$ and $L_3$ edges, lines denote fits proportional to $j$.
		(b) Magnetic spin and orbital moment per atom obtained in a sum rules analysis, with line fits.}
	\label{current-dep-fit}
\end{figure}

The current dependence of the AC-XMCD is best seen when integrating the respective values at the $L_3$ and $L_2$ edges. We show the result in Fig.~\ref{current-dep-fit}(a), together with a linear fit of the induced XMCD.
This finding suggests an out-of-plane reorientation of the Co magnetization proportional to the applied current density.
For a quantitative analysis of the magnetic moments we use the XMCD sum rules \cite{Thole1992,Carra1993,Chen1995}.
The resulting magnetic moments are plotted as a function of current density in Fig.~\ref{current-dep-fit}(b), together with their linear fits.
For the highest $j=3.78 \times 10^{6}$~\Acm we found a strong drop of the static magnetic moment to 20--25\% (not shown), meaning that we already heat the sample close to the Curie temperature.

The ratio of the AC-moments to the static moments gives us the out-of-plane deviation angle of the magnetization, $\alpha = 0.058^\circ$, for $j = 10^7$~\Acm assuming homogeneous current density.
We now use this value to determine the size of the dampinglike field $B_{DL}$ that is the origin of $M_z$.
Together with the sample's out-of-plane anisotropy of $B_A = 712$~mT, as determined in a separate harmonic Hall voltage measurement, we get $B_{DL} = B_A \sin(\alpha) = 0.72$~mT, for $j = 10^7$~\Acm.
This value is slightly lower than the $1.17$~mT found in harmonic Hall voltage measurements \cite{Avci2014} for the same current density and film thicknesses.
Further, we determine the spin-orbit torque efficiency to be \cite{Nguyen2016}
$$\xi_{DL} = \frac{2e}{\hbar j} M_s t_{Co} B_{DL} = 0.078$$
with about 15\% error margins from the determination of the moments, and
using $\mu_0 M_s = 1.8$~T for Co.
The effective spin Hall angle is often taken equal to the spin-orbit torque efficiency, $\theta^{eff}=\xi_{DL}$. Alternatively, assuming a drift-diffusion model of the spin Hall effect, the spin-orbit torque efficiency is corrected for the finite Pt thickness \cite{Liu2011}, giving
$$\theta^{eff} = \frac{\xi_{SH}}{1-1/\cosh(t_{Pt}/\lambda_{Pt})},$$
where $\lambda_{Pt}$ is the spin diffusion length of Pt, with reported values that range from 1 to 14~nm \cite{Sinova2015,Sagasta2016}. Finally we note that any fieldlike torque on the Co layer is not detected due to the chosen measurement geometry.
It has been shown previously that in Pt(6~nm)/Co(2.5~nm) the fieldlike torque is about one order of magnitude smaller than the dampinglike torque \cite{Avci2014}.
Similarly, we are insensitive to the Oersted field generated by the current through the sample, which is acting along $y$ as the $z$ component averages to zero.
An analytic calculation of the Oersted field \cite{Hayashi2014} gives an average value of 0.38~mT within the Co layer.

\section{Results on P\lowercase{t}/NM bilayers}

Numerous measurements on Pt/Co and other magnetic bilayers are present in the current literature, but investigations on Pt/NM are rare.
Choosing a nonmagnetic layer in contact with Pt has the advantage of avoiding the direct interaction that the accumulated spins would have with the Co magnetic moment.
This is apparent if one compares values of the spin diffusion length: whereas for a single Pt layer at room temperature up to 11~nm was found \cite{Stamm2017}, this length is reduced to values between 1.1 \cite{Avci2015} and 2~nm \cite{Sinova2015} if Pt is in contact with Co.
Direct XMCD studies at the Pt $L_3, L_2$ edges \cite{Grange1998} would circumvent this difficulty, but the Pt $L$ edges around $11.5$~keV photon energy have a lower cross section and reduced XMCD contrast compared to the 3\textit{d} transition metal edges, resulting in a sensitivity of $\approx 10^{-3}~\mu_B$ per Pt atom \cite{Geprags2012}. Moreover, given the large penetration depth of hard x-rays, one still faces the difficulty of the inhomogeneous spin distribution within the Pt layer: the accumulation on the top interface is balanced by the same accumulation of spins with opposite sign at the bottom interface \cite{Zhang2000}.
To avoid cancellation of the effect in the measurement, the employed technique would need to be sensitive to one layer only.
We therefore follow a different approach, in which we add a nonmagnetic (NM) indicator layer of Ti, Cr, or Cu on top of Pt.
We then perform highly sensitive XMCD measurements on the NM film of the Pt/NM bilayer under current injection.

\subsection{Pt/Ti}

\begin{figure}[tb]
	\includegraphics[scale=0.53]{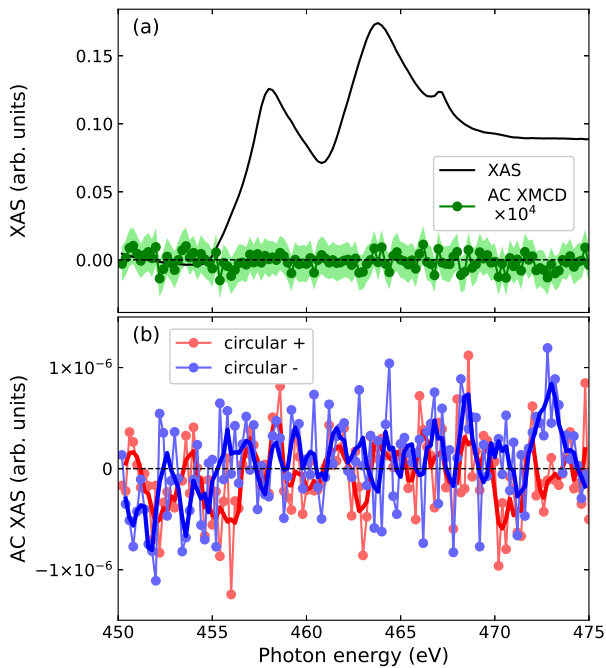}
	\caption{Pt/Ti bilayer under current injection: (a) XAS (black line) and current induced XMCD (green dots, standard error marked as a light-green band) magnified by the given factor. (b) current-induced change of the absorption, for circular positive and circular negative x-ray helicity.}
	\label{PtTi}	
\end{figure}

We inject a current of 62~mA through a 200~\mum wide strip of the Pt(6~nm)/Ti(6~nm) bilayer sample and record XAS and XMCD spectra shown in Fig.~\ref{PtTi}.
Because of the higher resistivity of the Ti layer, the current density in Pt will be increased from $j=2.6\times 10^6$~\Acm for homogeneous current distribution, to about $j=4.1\times 10^6$~\Acm.
The measurement was performed in the geometry drawn in Fig.~\ref{sample}(b).
The angle of 60$^\circ$ ensures the XMCD to be predominantly sensitive to the in-plane magnetic moment, transverse to the current direction, which corresponds to the direction of spins that accumulate due to the spin Hall effect \cite{Stamm2017}.

Even after averaging 56 consecutive energy scans, no dichroic signal could be detected in Ti: the XMCD signal in Fig.~\ref{PtTi}(a) fluctuates around the zero line, and the x-ray helicity resolved XAS curves, which give the XMCD when subtracted from each other, are almost identical within the noise limit, see Fig.~\ref{PtTi}(b).
To estimate the amount of magnetic moment in Ti, we integrated the XMCD signal around the $L_3$ and $L_2$ edges in a sum rule analysis, and found an upper limit for the induced magnetic moment of $5\times 10^{-6}~\mu_B$ per atom.

\subsection{Pt/Cr}

\begin{figure}[tb]
	\includegraphics[scale=0.53]{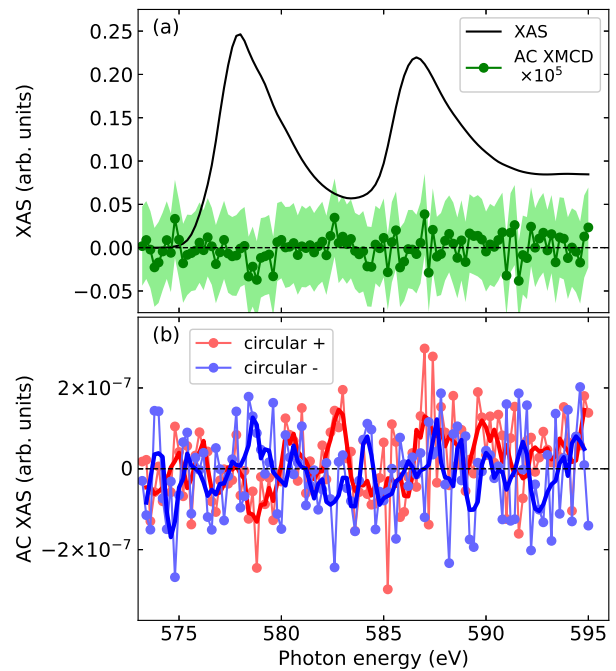}
	\caption{Pt/Cr bilayer under current injection. (a) and (b) as in Fig.~\ref{PtTi}.}
	\label{PtCr}
\end{figure}

Next we describe the results of injecting current into a 200~\mum wide strip of Pt(6~nm)/Cr(6~nm).
As was recently observed, Cr has a strong spin-orbit coupling leading to a sizable spin Hall angle of about half that of Pt \cite{Du2014}.
In the transmission measurement we however integrate the magnetic moment over the whole film thickness, leading to a cancellation of the magnetic signal should there exist a symmetric separation of spins inside the Cr layer.
Only an overall spin accumulation that presumably forms at the Pt/Cr interface will cause an XMCD signal in Cr.
Therefore we are sensitive to the spin-orbit effects in Pt acting on Cr, but not to the intrinsic effects in the Cr layer.

\begin{figure}[tb]
	\includegraphics[scale=0.53]{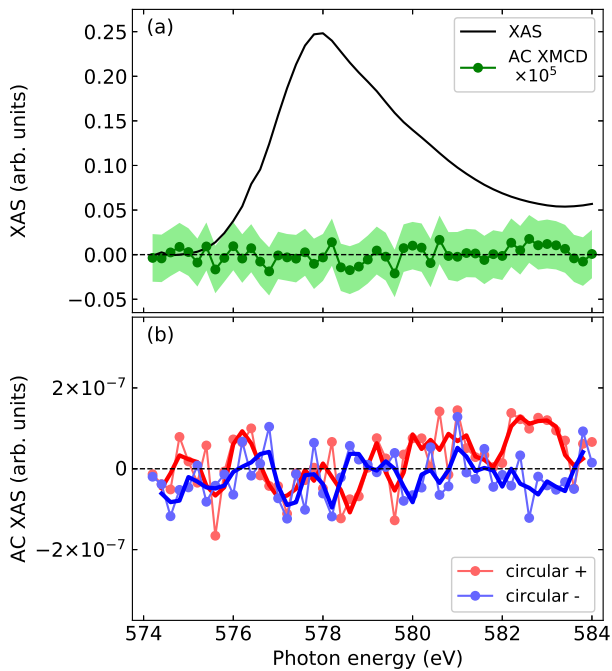}
	\caption{Pt/Cr measurement under current injection, focused around the Cr $L_3$ edge. (a) and (b) as in Fig.~\ref{PtTi}.}
	\label{PtCrL3}
\end{figure}

A current of 72~mA, yielding a current density of $j=3\times 10^6$~\Acm, was injected through Pt, and the sample was measured at an angle of 60$^\circ$, see Fig.~\ref{sample}(b).
A very sensitive measurement was achieved by averaging 144 absorption scans acquired during almost 30 h, the result is shown in Fig.~\ref{PtCr}.
A small hint of a possible spin signal is visible around the $L_3$ absorption line, as a dip around $E=579$~eV in Fig.~\ref{PtCr}(a), also visible as differences in the x-ray helicity resolved signals in Fig.~\ref{PtCr}(b).
We notice however that noise related fluctuations are of the same amplitude as the presumed spin signal.
To further improve the signal to noise ratio, we concentrate further measurements on the Cr $L_3$ edge in Fig.~\ref{PtCrL3}, the result of averaging 416 spectra.
While here we have only half the noise compared to the data shown in Fig.~\ref{PtCr}, a clear sign of an induced magnetic signal at the Cr $L_3$ edge is still elusive.
We therefore only give an upper limit of a possible spin signal, which amounts to $\approx 2\times 10^{-6}~\mu_B$ per Cr atom, as found in a sum rule integration.

\subsection{Pt/Cu}

\begin{figure}[!htb]
	\includegraphics[scale=0.53]{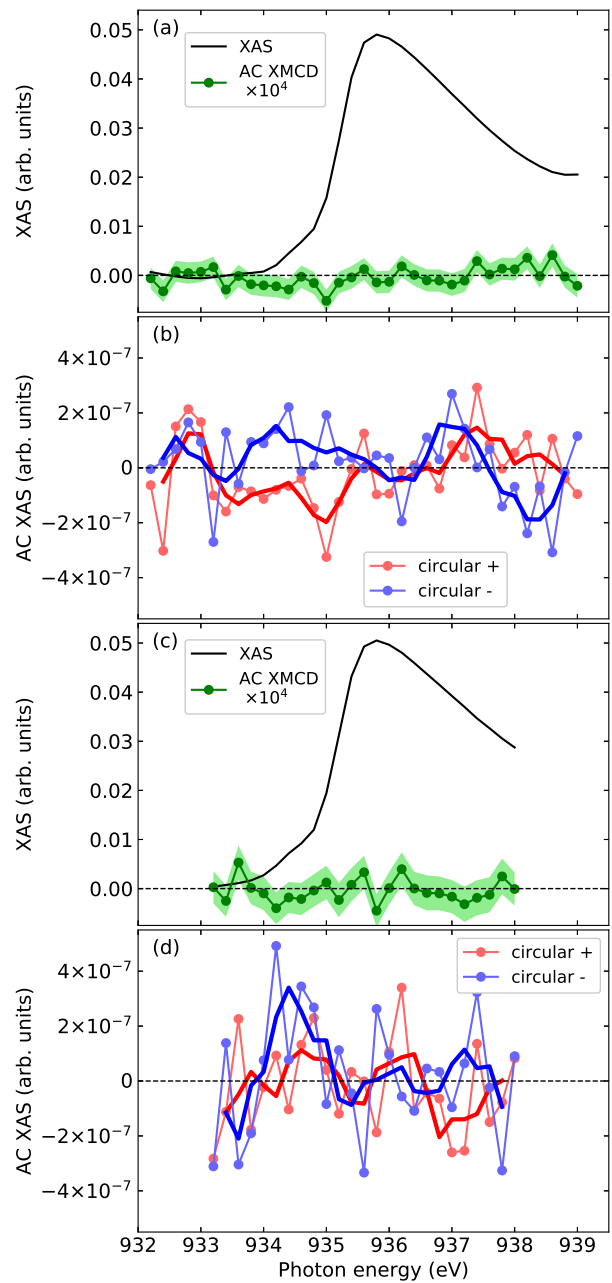}
	\caption{XAS and XMCD of Pt/Cu, measured at the Cu $L_3$ edge. 
		(a) Pt(10 nm)/Cu(10 nm) subjected to a current of 100~mA:		
		XAS (black line) and current induced XMCD (green dots) magnified by $10^4$, with XMCD standard error as a light-green band.
		(b) current induced change of the absorption, for circular positive and circular negative x-ray helicity. 
		(c), (d) Same graphs as (a) and (b), for Pt(20 nm)/Cu(10 nm) subjected to a current of 80~mA.}
	\label{PtCu}
\end{figure}

At last we turn our attention to Pt/Cu bilayers.
Cu is a particularly important element for spintronics applications, as it has one of the longest spin diffusion lengths.
The excellent conductivity of Cu however is a disadvantage for our current-in-plane geometry, as a considerable fraction of the current will actually flow in the Cu layer and thus does not contribute to spin accumulation in Pt.
Measurements have been performed on samples with Pt thicknesses of 6, 10, and 20~nm, each covered by 10~nm of Cu, and strips of reduced width of 100~\mum.
The results look qualitatively very similar, and we show in Fig.~\ref{PtCu} the result of the Pt(10 nm)/Cu(10 nm) sample.
The 100~mA current resulted in a current density in the Pt layer of $j=2.6 \times 10^{6}$~\Acm after accounting for the shunting through Cu, using thin film resistivity values \cite{Dutta2017}.

A very small XMCD signal is visible in Fig.~\ref{PtCu}(a), also seen as opening of the two curves with opposite x-ray helicity as plotted in Fig.~\ref{PtCu}(b).
A comparison to the XMCD sum rule analysis of Cu polarized in the proximity of Co \cite{Samant1994} allows us to estimate a magnetic moment of $\approx 1.5 \times 10^{-6}~\mu_B$ on Pt(10~nm)/Cu(10~nm), with an error margin of the same size.
Similar results come from measurements on a Pt(20~nm)/Cu(10~nm) sample shown in Fig.~\ref{PtCu}(c,d). 
Without judging the significance of the signal, we conclude that there may be signs of spins injected from the accumulation in Pt, but their moments average in the Cu layer to less than $\approx 3 \times 10^{-6}~\mu_B$ per atom.

\section{Discussion and summary}

Summarizing our measurements on Pt/Co under current injection, we found that the  magnetization of Co reorients itself from purely in-plane into a new equilibrium position with a small perpendicular component.
This process is described by a dampinglike field, a consequence of the current-induced accumulation of spins at the Pt/Co interface.
The quantitative analysis of our XMCD measurements reveals a spin Hall angle of $\theta^{eff} = 0.08$.
In addition, our XMCD analysis shows a constant $m_L / m_S$ ratio in Co that is independent of the injected current density.

Next we exchanged Co with a nonmagnetic 3\textit{d} transition metal and measure with high accuracy the amount of spins that emerge in Ti, Cr, or Cu.
Any signal of accumulated spins is however so small that we can only give upper limits of spin accumulation, with the exception of Cu which may exhibit a moment of $\approx 1.5 \times 10^{-6}~\mu_B$ although with error bars of the same size.
A similar conclusion was reported in an attempt to measure the injected spins in a lateral spin valve, for which the authors could not reveal magnetic contrast on the nonmagnetic Cu electrodes but gave an upper limit of $10^{-4} \mu_B$ XMCD contrast \cite{Mosendz2009}.
Only in the measurements reported on a Co/Cu nanopillar, a small but measurable spin moment of $3 \times 10^{-5}~\mu_B$ was detected \cite{Kukreja2015}.
In the nanopillar the current is forced to flow perpendicular, from Co into Cu, and thus the spins are transported into Cu with very high efficiency.

We now compare the result on Pt/Cu presented here with our previous magneto-optical measurement of the spin Hall effect induced spin accumulation in a single Pt layer \cite{Stamm2017}.
At our current density, one expects an accumulation of $1.3 \times 10^{-5}~\mu_B$ at the surface of Pt, for thickness $t_{Pt} \gg \lambda_{Pt}=11~\mathrm{nm}$.
For the 10~nm Pt thickness used here, this value would be reduced to $\approx 4 \times 10^{-6}~\mu_B$ using the correction for finite Pt thickness \cite{Sinova2015,Liu2011} but with the spin diffusion length of a single Pt layer \cite{Stamm2017} of $\lambda_{Pt}=11~\mathrm{nm}$.
Experimentally we determined the moment in Cu to be lower than $\approx 3 \times 10^{-6}~\mu_B$ for 10~nm Pt/Cu sample, which confirms that we are in the range of detecting the magnetic moment.
An unknown factor however is the role of the Pt/Cu interface, which may introduce spin loss due to electron scattering, leading to a further reduction of detectable spins in the Cu layer.
As we see from the correction for finite Pt thickness, a thicker Pt layer would help increasing the spin accumulation.
We additionally measured a Pt(20~nm)/Cu(10~nm) sample with twice the Pt thickness, and would expect roughly a two-fold increase of the AC-XMCD signal.
However, the 20~nm Pt/Cu has a lower current density in Pt of $1.8 \times 10^6$~\Acm.
Overall we would expect the AC-XMCD to be similar in size for the two Pt/Cu samples, which is indeed found comparing panels a and c in Fig.~\ref{PtCu}.

Finally we estimate the influence of the Oersted field, which lies along the same axis as the presumed accumulated moment in the Pt/NM current strips.
The calculation for Pt(10~nm)/Cu(10~nm) gives a value of $B_{Oe} = 0.09$~mT, which leads to an induced magnetic moment $M = \chi B / \mu_0$, with
$\chi = -9.63 \times 10^{-6}$
being the magnetic susceptibility of Cu \cite{Schenck1996}.
In Pt/Cu, the Oersted field will thus induce a magnetic moment corresponding to $\approx10^{-9}~\mu_B$ per atom, well below our experimental sensitivity and far lower than the presumed spin accumulation signal.
Analogous numbers are found for Pt/Ti and Pt/Cr, and we thus conclude that contributions from the Oersted field can be neglected in Pt/NM bilayers.

\begin{acknowledgments}
	
	We thank C.O.\ Avci for fruitful discussions and T.\ B{\"a}hler, L.\ Debenjak, and P.\ Schifferle for technical assistance.
	We acknowledge the Paul Scherrer Institut, Villigen, Switzerland for provision of synchrotron radiation beamtime at the SIM beamline of the Swiss Light Source, and  the Swiss National Science Foundation for financial support through Grants No.\ 200021-153404 and No.\ 200020-172775.
		
\end{acknowledgments}

\appendix*

\section{Effective current density in Pt layer}

When injecting current into a Pt/NM bilayer sample, both layers will conduct a fraction of the current according to their electrical resistivity and thickness.
Here we estimate the individual current density inside each layer of the Pt/Co sample by assuming two resistors in parallel, $R_{Pt}$ and $R_{Co}$.
The voltage across each resistor is identical, $U_{Pt} = U_{Co} = R_{Pt} I_{Pt} = R_{Co} I_{Co}$, whereas the total current is given by the sum of the currents through each layer, $I = I_{Pt} + I_{Co}$.
We get the current ratio as the inverse of the resistance ratio
$$r = \frac{I_{Pt}}{I_{Co}} = \frac{R_{Co}}{R_{Pt}} = \frac{\rho_{Co}}{\rho_{Pt}} \frac{t_{Pt}}{t_{Co}} ,$$
using the definition of the resistance $R_i = \rho_i l / (t_i w)$, with identical strip length $l$ and width $w$ for both layers, and independent resistivity $\rho_i$ and thickness $t_i$ values.
For Pt(6~nm)/Co(2.5~nm) we use values found in a separate four-point measurement on Hall bar structures,
$\rho_{Pt} = 255~\mathrm{n\Omega\,m} \qquad \mathrm{and} \qquad \rho_{Co} = 600~\mathrm{n\Omega\,m}$,
resulting in $r=5.65$.

The fraction of current in each layer is given by
$$\frac{I_{Pt}}{I} = \frac{r}{1+r} = 0.85 \; \mathrm{and} \;
\frac{I_{Co}}{I} = \frac{1}{1+r} = 0.15 .$$
The relative current density in Pt then becomes
$$\frac{j_{Pt}}{j} = \frac{I_{Pt}}{I} \frac{t_{Pt}+t_{Co}}{t_{Pt}} = \frac{r}{1+r}  \frac{8.5~\mathrm{nm}}{6~\mathrm{nm}} = 1.20 ,$$
or a $20\%$ increase compared to the current density that was calculated assuming a homogeneous distribution.

\end{document}